# Dynamics of quartz tuning fork force sensors used in scanning probe microscopy

A Castellanos-Gomez[1], N Agraït[1,2,3] and G Rubio-Bollinger[1,2]

[1] Departamento de Física de la Materia Condensada (C–III).
Universidad Autónoma de Madrid, Campus de Cantoblanco, 28049 Madrid, Spain.
[2] Instituto Universitario de Ciencia de Materiales "Nicolás Cabrera".
[3] Instituto Madrileño de Estudios Avanzados en Nanociencia
IMDEA-Nanociencia, 28049 Madrid, Spain

E-mail: gabino.rubio@uam.es

We have performed an experimental characterization of the dynamics of oscillating quartz tuning forks which are being increasingly used in scanning probe microscopy as force sensors. We show that tuning forks can be described as a system of coupled oscillators. Nevertheless, this description requires the knowledge of the elastic coupling constant between the prongs of the tuning fork, which has not yet been measured. Therefore tuning forks have been usually described within the single oscillator or the weakly coupled oscillators approximation that neglects the coupling between the prongs. We propose three different procedures to measure the elastic coupling constant: an opto-mechanical method, a variation of the Cleveland method and a thermal noise based method. We find that the coupling between the quartz tuning fork prongs has a strong influence on the dynamics and the measured motion is in remarkable agreement with a simple model of coupled harmonic oscillators. The precise determination of the elastic coupling between the prongs of a tuning fork allows to obtain a quantitative relation between the resonance frequency shift and the force gradient acting at the free end of a tuning fork prong.

1. **Introduction**

Quartz tuning forks (TFs) have been widely used as force sensors in scanning probe microscopes (SPMs) to image and to manipulate matter at the nanoscale [1-6]. Miniaturized quartz TFs are mass produced as the time-base in the watch industry. To convert one of these miniaturized TFs into an SPM force sensor a sharp tip is attached to one its prongs. When a force gradient is acting on the tip, the resonance frequency is shifted making these force sensors useful for SPMs. The readout of these sensors is based on the native piezoelectric effect of quartz which yields an electrical current proportional to the deformation of the TF prongs. Therefore optical setups are not needed making easy the implementation of TF sensors in SPMs in ultrahigh vacuum. Moreover only one extra electrical connection is needed to supplement a scanning tunnelling microscope (STM) with a TF sensor, and the low power dissipation assures low temperature compatibility [7-9]. Unlike conventional microfabricated cantilevers, TF sensors are very stiff (elastic constant $k = 10^3 - 10^4 \, \text{Nm}^{-1}$) making possible to achieve stable small oscillation amplitudes without the tip jumping to contact at very small tip to sample distances [10, 11]. This small oscillation amplitude combined with the extremely high quality factor $Q$ of TFs enables the detection of small frequency shifts of the resonance frequency, allowing for atomic resolution imaging [12, 13] and high sensitivity measurement of atomic scale forces [5, 6].

Due to the high stiffness and $Q$ factor of TF sensors, it is very convenient to use the frequency modulation (FM) scheme in SPMs based on TF sensors. In this scheme the TF is driven at its resonance frequency by means of a *phase locked loop* (PLL) circuit and the frequency shift is measured. This frequency shift is related with the force gradient acting between the TF tip and the surface of the studied sample. Although the use of TF sensors with frequency modulation (FM) technique is mature enough, it is still hard to quantitatively obtain the tip-sample force gradient from the resonance frequency shift. The reason is that the formalism developed for conventional cantilevers is not strictly valid for TFs. In the Q-Plus configuration proposed by Giessibl [14], where a TF is turned into a quartz cantilever by firmly gluing one prong to a massive holder, the force gradient can be extracted from the frequency shift as in conventional cantilevers [15]. Nevertheless the $Q$ factor in the Q-Plus configuration can be highly dependent on the way the fixed prong is glued.

J. Rychen [16] already noticed that any asymmetry in a TF results in additional damping reducing the $Q$ factor. Therefore using a balanced TF with the two free prongs takes advantage of a higher $Q$ factor. To relate the force gradient acting between the tip and the surface to the resulting frequency shift, the TF effective elastic constant has to be determined. G.H. Simon *et al.* [17] found that the effective elastic constant of a TF it is not straightforwardly related with the elastic constant of a TF prong considered as a cantilever. This is due to the strong effect of the coupling between the TF prongs on its dynamics.

In this context we present an experimental study of the double prong TF dynamics. We find that the experimentally observed TF dynamics and the thermal noise spectra are in remarkable agreement with a coupled oscillators model. We have developed three different calibration methods that allow obtaining the effective elastic constant of TF sensors and the elastic constant of the coupling between the prongs. Thus the force gradient acting between the tip and the sample can be obtained from the measured frequency shift in the resonance frequency. We also present a procedure to counteract the mass unbalance due to the attachment of a tip to TF sensors.

## 2. Two coupled oscillators model

TF sensors have been widely modelled in the literature by a single harmonic oscillator [18, 3, 19, 20]. But we have found that the experimentally observed TF motion does not match with the one expected for a single harmonic oscillator. The reason is that TFs behave as a pair of coupled cantilevers and thus their dynamics is strongly dependent on the coupling.

Let us to initially model the prongs of the TF as two identical harmonic oscillators (1 and 2) with effective masses $m$ and elastic constants $k$. The coupling between the prongs is modelled by a spring with elastic constant $k_c$. The motions of the masses 1 and 2 ($x_1$ and $x_2$) represent the deflections of the free ends of the prongs 1 and 2 of the TF. The equations of motion are

$$\left. \begin{array}{l} m\ddot{x}_1(t) + (k+k_c)x_1(t) - k_c x_2(t) = 0 \\ m\ddot{x}_2(t) + (k+k_c)x_2(t) - k_c x_1(t) = 0 \end{array} \right\}. \quad (1)$$

Within the harmonic approximation the eigenmodes that solve this equation system are: one in which the masses oscillate in-phase with the same amplitudes and other in which the masses oscillate in anti-phase with the same amplitudes (see figure1a). The coupling between the two harmonic oscillators breaks the degeneracy of the uncoupled identical oscillators and thus the eigenfrequencies are

$$\left. \begin{array}{l} f_0^{\text{in-phase}} = \dfrac{1}{2\pi}\sqrt{\dfrac{k}{m}} \\ f_0^{\text{anti-phase}} = \dfrac{1}{2\pi}\sqrt{\dfrac{k+2k_c}{m}} \end{array} \right\}, \quad (2)$$

where the superscript labels the eigenmode and the subscript 0 specifies that the two oscillators are identical as it would be the case of a perfectly balanced TF. From equation (2) the elastic constant of the coupling $k_c$ can be easily expressed in terms of the elastic constant of one prong $k$ and the eigenfrequencies of the two identical coupled oscillators $f_0^{\text{in-phase}}$ and $f_0^{\text{anti-phase}}$:

$$k_c = \dfrac{k}{2}\left[\left(\dfrac{f_0^{\text{anti-phase}}}{f_0^{\text{in-phase}}}\right)^2 - 1\right]. \quad (3)$$

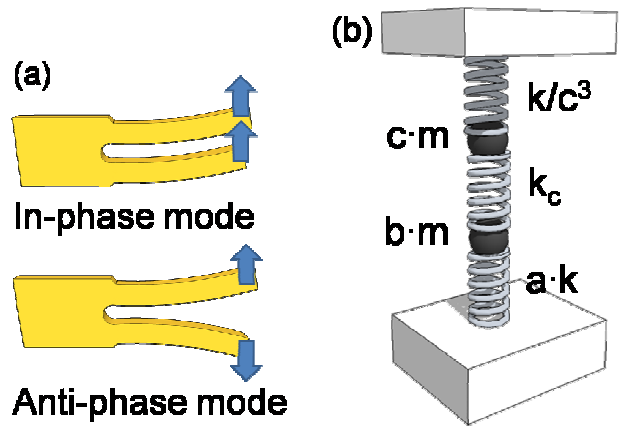

Figure 1. (a) Illustration of the in-phase and anti-phase eigenmodes of a TF. (b) Schematic diagram of the proposed coupled harmonic oscillators model where each prong is modelled as a mass and a spring and a central spring with elastic constant $k_c$ is added to model the coupling between the prongs. The parameter $a = 1 + \Delta k / k$ takes into account the effect of an external force gradient $\Delta k$ acting on one prong, the parameter $b = 1 + \Delta m / m$ takes into account the effect of an extra mass $\Delta m$ attached to one prong and $c = L_{\text{prong 2}} / L_{\text{prong 1}}$ takes into account that the length of the prongs $L$ can be different.

A more realistic model should take into account that the oscillators are not identical (see figure 1b). First, the prongs could be slightly different in shape and therefore their masses and elastic constants would be different. Here we have also taken into account that the length of the prongs $L$ can be different. Note that if the prongs are considered as cantilevers the mass of one prong $m$ is proportional to its length $L$ and the elastic constant of the prong $k$ is proportional to $L^{-3}$. Second, usually one prong is mass loaded with a tip increasing the mass of this prong $m$ by an amount $\Delta m$. And last but not least, the force gradient $\Delta k$ between the tip and the surface acts only on one of the prongs, effectively unbalancing the TF. Taking into account these considerations the equations of motion of the masses 1 and 2 are given by:

$$\left. \begin{array}{l} bm\ddot{x}_1(t) + (ak + k_c)x_1(t) - k_c x_2(t) = 0 \\ cm\ddot{x}_2(t) + \left(\dfrac{k}{c^3} + k_c\right)x_2(t) - k_c x_1(t) = 0 \end{array} \right\}, \quad (4)$$

where $a = 1 + \Delta k / k$, $b = 1 + \Delta m / m$ and $c = L_2 / L_1$ is the length ratio of the prongs. This equation system can be rewritten using equation (2) in terms of $x_1(t)$, $x_1(t)$, *a*, *b*, *c* and the identical coupled oscillator eigenfrequencies: $f_0^{\text{in-phase}}$ and $f_0^{\text{anti-phase}}$. This is important because frequencies can be much more accurately measured than spring constants or masses. From this equation system it is straightforward to obtain the relation between the TF frequency shift and the force gradient $\Delta k$ acting on one prong. Considering the case in which both oscillators have identical masses $m$ and lengths $L$ ($b = 1$ and $c = 1$) and only a small force gradient $\Delta k \ll k$ is present, the frequency shift $\Delta f^{\text{anti-phase}} = f^{\text{anti-phase}} - f_0^{\text{anti-phase}}$ is

$$\frac{f^{\text{anti-phase}} - f_0^{\text{anti-phase}}}{f_0^{\text{anti-phase}}} \simeq \frac{1}{2}\frac{\Delta k}{2(k + 2k_c)}, \qquad (5)$$

with $f^{\text{anti-phase}}$ the anti-phase eigenfrequency for the unbalanced TF due to the force gradient. Only the anti-phase mode frequency shift is shown because this mode is the one commonly used for sensing applications. The reason is that the TF electrodes design nulls out the net current from the prongs for the in-phase oscillation. Equation (5) shows that the frequency shift $\Delta f^{\text{anti-phase}}$ is proportional to the force gradient $\Delta k$ like in a cantilever but with an effective elastic constant $k_{\text{eff}} = 2(k + 2k_c)$ which is at least twice that of a single oscillator and is strongly dependent on the coupling between the prongs. Using equation (3) the effective elastic constant $k_{\text{eff}}$ can be written in terms of the elastic constant of one prong ($k$) and the eigenfrequencies of the two identical coupled oscillators $f_0^{\text{in-phase}}$ and $f_0^{\text{anti-phase}}$

$$k_{\text{eff}} = 2k\left(\frac{f_0^{\text{anti-phase}}}{f_0^{\text{in-phase}}}\right)^2. \qquad (6)$$

A common method to measure the effective elastic constant of microcantilevers is the so called Cleveland method [21] that consists in measuring the resonant frequency of the cantilever before and after adding small end masses. The coupled harmonic oscillators model can be used to adapt the Cleveland method to TF sensors. Considering the case in which there is no force gradient applied ($a = 1$), the length of the prongs is identical ($c = 1$) and a small mass $\Delta m \ll m$ is added to the end of one prong, the relation between the added mass $\Delta m$ and the resonance frequency $f^{\text{anti-phase}}$ is

$$\Delta m \simeq \frac{k_{\text{eff}}}{\left(2\pi f^{\text{anti-phase}}\right)^2} - 2m. \qquad (7)$$

This expression resembles equation (4) in Ref. [21] $\Delta m = k \cdot (2\pi f)^{-2} - m$, valid for a single oscillator with effective mass $m$ and elastic constant $k$ but with the TF effective elastic constant $k_{\text{eff}}$ instead of $k$ and a factor 2 in the effective mass $m$, which is related to the fact that in a TF both prongs are moving.

Another procedure widely used to determine the effective elastic constant of microcantilevers is the measurement of their thermal noise [22, 23]. This technique can be extended to TFs within this coupled harmonic oscillators model. The hamiltonian $H$ of the system can be written in terms of the normal coordinates $z^{\text{in-phase}} = x_1 + x_2$ and $z^{\text{anti-phase}} = x_1 - x_2$. Assuming that the elastic constants and the masses of both oscillators are identical ($a = 1$, $b = 1$ and $c = 1$) the hamiltonian $H$ is

$$H = \frac{\left(\dot z^{\text{anti-phase}}\right)^2}{4m} + \frac{\left(\dot z^{\text{in-phase}}\right)^2}{4m} + \frac{k}{4}\left(z^{\text{in-phase}}\right)^2 + \frac{1}{2}\left(\frac{k}{2} + k_c\right)\left(z^{\text{anti-phase}}\right)^2. \qquad (8)$$

By virtue of the energy equipartition theorem and taking into account that $x_1 = -x_2 = x$ in the anti-phase mode, the effective elastic constant $k_{\text{eff}}$ is

$$k_{\text{eff}} = \frac{k_B T}{\langle x^2 \rangle}, \qquad (9)$$

where $k_B$ is the Boltzmann constant and $T$ is the absolute temperature.

Summarizing, the TF effective elastic constant $k_{\text{eff}}$ can be obtained in three alternative ways:

1. Following expression (6), measuring the eigenfrequencies ratio $f_0^{\text{anti-phase}} / f_0^{\text{in-phase}}$ and using the elastic constant k calculated from prong's dimensions.
2. By a variation of the method developed by Cleveland et al. [21] measuring the change in the anti-phase eigenfrequency $f^{\text{anti-phase}}$ while one of the prongs is mass loaded with $\Delta m$.
3. From the thermal noise measurement using formula (9).

Once the effective elastic constant $k_{\text{eff}}$ is obtained the TF sensor sensitivity $\alpha \equiv \Delta f^{\text{anti-phase}} / \Delta k$ can be obtained:

$$\alpha = \frac{f_0^{\text{anti-phase}}}{2 k_{\text{eff}}}. \qquad (10)$$

We have found these three approaches to obtain the TF sensitivity yield consistent values for the sensitivity. Additionally, if the elastic constant of the prongs (*k*) is known, the elastic coupling constant $k_c$ can be easily obtained from the effective elastic constant $k_{\text{eff}}$

$$k_c = \frac{k_{\text{eff}} - 2k}{4} \qquad (11)$$

### 3. Experimental details

The calibration methods based on the model presented in the previous section have been carried out with the setup sketched in figure 2a. A TF is excited mechanically by a dither piezo while it is inspected under an optical microscope (Nikon Eclipse LV-100). The combination of mechanical excitation and optical inspection allows also the measurement of the in-phase mode which, due to the TF electrodes design, can not be self-excited or detected electrically. During optical inspection the TF piezoelectric current is measured using a current to voltage converter and detected in phase with a lock in amplifier (SR830 DPS Stanford).

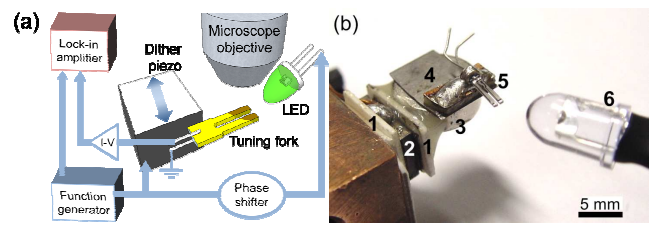

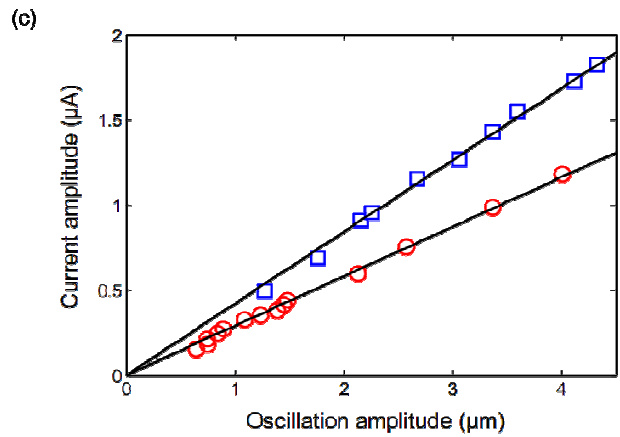

Figure 2. (a) Experimental setup for the optical characterization of the TF dynamics. (b) Photograph of the experimental setup: (1) alumina plates, (2) dither piezo with a calibration of 3 nm/V, (3) magnet, (4) TF steel holder, (5) TF and (6) high intensity LED. This setup is mounted under a long working distance objective of a Nikon Eclipse LV-100 optical microscope. (c) Measured piezoelectric current amplitude against anti-phase oscillation amplitude for two different TFs, TF-A (red circles) and TF-B (blue squares). The current amplitude and the oscillation amplitude are proportional.

Illuminating the TF with a light emitting diode (LED) modulated at the dither frequency (i.e. stroboscopic illumination) provides a very convenient optical in-phase motion detection. If the phase between the illumination and the excitation is shifted the whole eigenmode motion can be explored. In-phase and anti-phase modes can be easily identified by this procedure. Moreover by driving the LED with a frequency slightly shifted with respect to the dither frequency, the TF oscillation can be filmed with a regular CCD camera (see online multimedia attachment). In order to optically measure the oscillation amplitude it is convenient to illuminate the TF at twice the dither frequency. In this way the resulting image is a superposition of two instants of the oscillation phase shifted 180º. Adjusting the phase shift between illumination and excitation both extremals of the oscillation can be simultaneously observed. The relation between piezoelectric current and dither voltage is linear for TF oscillation amplitudes from 1Å to several µm. Moreover this oscillation can be optically detected for amplitudes larger than 0.5 µm as shown in figure2b where the oscillation amplitude is plotted against the piezoelectric current amplitude for two different TFs models[1] (TF-A and TF-B hereafter).

### 4. Validation of the coupled oscillators model

There is a change in the anti-phase eigenfrequency $f^{\text{anti-phase}}$ when the length of one prong *L* is reduced by mechanical cleavage. The coupled harmonic oscillators model presented in section 2 can account for this change in the anti-phase eigenfrequency $f^{\text{anti-phase}}$. Assuming that TF prongs are not mass loaded ($b=1$) and there is no force gradient applied ($a=1$) the TF dynamics only depends on the length ratio of the prongs $c = L_2 / L_1$ and the eigenfrequencies of the perfectly balanced TF: $f_0^{\text{anti-phase}}$ and $f_0^{\text{in-phase}}$ (see equation 4). Therefore, once the eigenfrequencies of the balanced TF are measured the anti-phase eigenfrequency

---

[1] We have purchased these TFs from Digikey. Digikey part number: SER3203-ND and SE3301-ND.

$f^{\text{anti-phase}}$ can be calculated as a function of the prongs length ratio $c$ by solving equation (4). As the length ratio of the prongs $c$ is easily measurable and no adjustable parameters are needed to calculate the anti-phase eigenfrequency $f^{\text{anti-phase}}$ against the length ratio, the comparison of the measured and the calculated relation provides a convenient method to check the validity of the coupled oscillators model. Figure 3 shows the experimental results (symbols) for a TF-A and a TF-B compared with the result obtained from the coupled harmonic oscillators model (broken lines). The excellent agreement validates the coupled oscillators model for TFs A and B. We have further checked that this model starts to fail for TFs which are poorly fixed at their base, causing the TF to oscillate as a whole for the in-phase eigenmode[2].

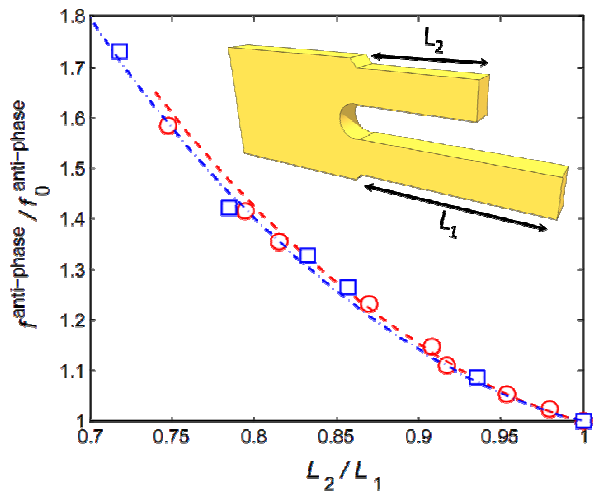

**Figure 3.** Dependence of the anti-phase eigenfrequency with the length ratio of the prongs measured for TF-A (red circles) and TF-B (blue squares). The calculated dependence within the coupled harmonic oscillators model is also plotted (red dashed line and blue dotted line for the TFs A and B respectively). Note that for the calculation only the experimental eigenfrequencies of the balanced TF are needed without any adjustable parameters.

### 5. Measurement of the effective elastic constant

In the following subsections the TF effective elastic constant $k_{\text{eff}}$ is measured using the three different approaches proposed in section 2.

*5.1. Opto-mechanical method*

First the in-phase $f_0^{\text{in-phase}}$ and anti-phase $f_0^{\text{anti-phase}}$ eigenfrequencies are measured for a bare TF using the experimental setup described in section 3. Then the dimensions of the prongs have been measured with an optical microscope. The elastic constant of one prong $k$ is calculated using the formula for a rectangular cross-section cantilever $k = 0.25EW\left(TL^{-1}\right)^3$ with $E$ the Young's modulus of quartz (78.7 GPa), $T$ the thickness, $W$ the width and $L$ the length of the TF prongs. With the eigenfrequencies of the bare TF and the prong's elastic constant $k$, the elastic constant of the coupling between the prongs $k_c$ and the effective elastic constant $k_{eff}$ have been obtained from equations (3) and (6) respectively. The main source of error of this calibration method is due to the strong dependence of the prong's elastic constant $k$ with its geometrical dimensions. Commercially available TFs have prongs whose shape is not exactly that of a rectangular cross-section cantilever. Consequently, the accuracy of this method could be improved by using a procedure independent of the TF geometrical dimensions to determine the prong's elastic constant $k$.

---

[2] For example the 'Fox NC26LF-327' TFs manufactured by Fox electronics.

Table 1. Eigenfrequencies and geometrical dimensions measured for a TF-A and a TF-B using the experimental setup described in section 3. The elastic constant of one prong $k$ has been calculated from the prong's geometrical dimensions. The coupling elastic constant $k_c$ and the effective elastic constant $k_{eff}$ have been obtained from equations (3) and (6) respectively. Note that the effective elastic constant $k_{eff}$ is underestimated by at least 20-35% if the coupling elastic constant $k_c$ is neglected, i.e. within the weakly coupled oscillators approximation ($k_{eff} \simeq 2k$).

|  | TF-A | TF-B |
|---|---|---|
| $f_0^{\text{in-phase}}$ (Hz) | 18255 | 27800 |
| $f_0^{\text{anti-phase}}$ (Hz) | 20000 | 32766 |
| L (μm) | $3200 \pm 30$ | $2500 \pm 30$ |
| T (μm) | $235 \pm 2$ | $235 \pm 2$ |
| W (μm) | $125 \pm 2$ | $100 \pm 2$ |
| $k$ (N/m) | $974 \pm 40$ | $1634 \pm 79$ |
| $k_c$ (N/m) | $98 \pm 4$ | $318 \pm 15$ |
| $k_{eff}$ (N/m) | $2338 \pm 96$ | $4540 \pm 220$ |

Table 1 summarizes the relevant parameters of the dynamics obtained using the opto-mechanical calibration of a TF-A and a TF-B. Two interesting outcomes can be extracted from these measured values. First, within the commonly used single oscillator approximation [18, 3, 19, 20] ($k_{eff} \simeq k$) the effective elastic constant $k_{eff}$ is underestimated by about 140-180%. And second, if the elastic constant of the coupling $k_c$ is neglected, i.e. within the weakly coupled oscillators approximation [18, 24] ($k_{eff} \simeq 2k$), the effective elastic constant $k_{eff}$ is underestimated by about 20-35%. Therefore the previously used single oscillator or weakly coupled oscillators approximations are inaccurate for commercially available TFs. However TF dynamics are faithfully described within the coupled oscillators model making possible the use of TF sensors in quantitative SPM applications.

### 5.2. Cleveland method variation

The second procedure to obtain the effective elastic constant $k_{eff}$ is a variation of the method developed by Cleveland *et al.* [21] for cantilever calibration. The anti-phase eigenfrequency $f^{\text{anti-phase}}$ is measured while one of the prongs is mass loaded at its end. From expression (7) the relation between this mass load $\Delta m$ and the inverse of the anti-phase eigenfrequency squared $\left(f^{\text{anti-phase}}\right)^{-2}$ is linear in the limit of small added mass $\Delta m \ll m$. The TF effective elastic constant $k_{\text{eff}}$ and the prong's effective mass $m$ can be extracted from the slope and the y-axis interception of a linear fit respectively.

We have attached small test masses[3] to the end of one prong of TFs A and B. The test masses naturally stick to the prong due to the presence of a small amount of flux covering the masses. We have found unnecessary to solder the masses to the TF electrodes because the resonance frequency and the $Q$ factor do not change appreciably after soldering them. Thus test masses can be removed after the calibration making this procedure non destructive. Once the sphericity of the particles is checked under the optical microscope, their masses are determined by measuring their diameter using the density of the bulk material as it was done in ref. [21]. We have estimated that the flux increases the mass load in less than 1%. By using a 50x long working distance objective an uncertainty in the determination of the mass load of less than 10% can be achieved.

Figure 4 shows the expected linear relationship between the added mass $\Delta m$ and $\left(f^{\text{anti-phase}}\right)^{-2}$ obtained for TFs A and B. By means of a linear fit to these data (solid black lines) the effective spring constants $2285 \pm 52$ Nm$^{-1}$ and $4505 \pm 234$ Nm$^{-1}$ are obtained for the TF-A and the TF-B respectively. Their corresponding effective masses are $(7.2 \pm 0.2) \times 10^{-8}$ kg and $(5.3 \pm 0.3) \times 10^{-8}$ kg. Additionally, the elastic coupling constant $k_c$ has been obtained from equation (11) and the elastic constant of the prongs $k$ obtained in previous section. The resulting values are $85 \pm 15$ Nm$^{-1}$ and $309 \pm 43$ Nm$^{-1}$ for the TFs A and B respectively.

We have used the coupled oscillators model to study the change of the resonance frequency when one prong is mass loaded. Assuming that there is no force gradient applied ($a=1$) and the length of the prongs is identical ($c=1$), the TF dynamics only depends on the eigenfrequencies of the balanced TF ($f_0^{\text{in-phase}}$ and $f_0^{\text{anti-phase}}$) and the parameter $b = 1 + \Delta m/m$ (see equation 4). Thus, the eigenfrequencies ($f_0^{\text{in-phase}}$ and $f_0^{\text{anti-phase}}$) and the effective mass $m$ have to be determined in order to obtain the relation between the added mass $\Delta m$ and $\left(f^{\text{anti-phase}}\right)^{-2}$. The eigenfrequencies ($f_0^{\text{in-phase}}$ and $f_0^{\text{anti-phase}}$) have been measured with the experimental setup described in section 3 and the effective mass $m$ has been obtained from a linear fit to the experimental data. The relation $\Delta m$ against $\left(f^{\text{anti-phase}}\right)^{-2}$ obtained from the model is compared with the experimental results in figure 4. A simple harmonic oscillator model can not account for a non-linear behaviour of the relation $\Delta m$ against $\left(f^{\text{anti-phase}}\right)^{-2}$ observed experimentally for sufficient large values of mass load $\Delta m$ (inset in figure 4). However, the relation calculated from the coupled oscillators model remarkably matches the experimental data.

---

[3] 15-45 µm diameter spheres extracted from SN62 MP218 solder paste.

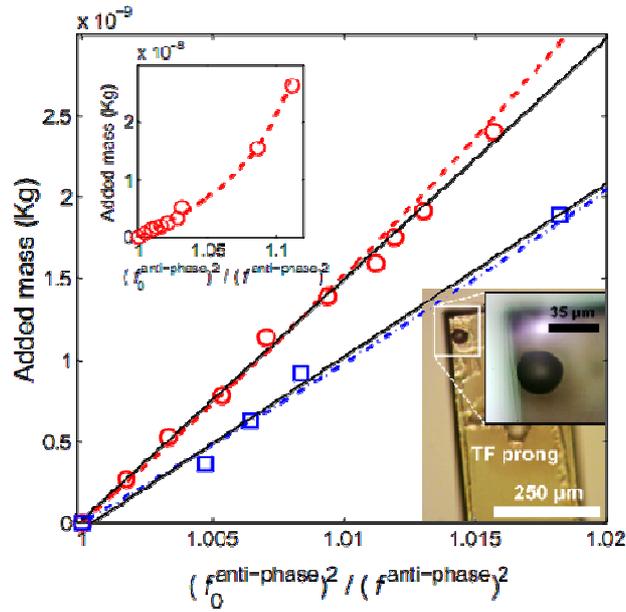

Figure 4. Plot of the added mass to the end of one prong $\Delta m$ against $\left(f^{\text{anti-phase}}\right)^{-2}$ for TF-A (red circles) and TF-B (blue squares). A linear fit (black solid lines) of the data gives the effective spring constants $2287 \pm 52$ Nm$^{-1}$ and $4505 \pm 234$ Nm$^{-1}$ for the TF-A and the TF-B respectively. The relation $\Delta m$ against $\left(f^{\text{anti-phase}}\right)^{-2}$ has been calculated within the coupled oscillators model for TF-A (red dashed line) and the TF-B (blue dashed-dotted line). The inset at the top shows the non linear behaviour of $\Delta m$ against $\left(f^{\text{anti-phase}}\right)^{-2}$ measured for a TF-A with large mass loads (symbols). The calculated result (dashed line) it is also shown. The inset at the bottom shows an optical micrograph of a test mass attached to the end of one TF prong.

### 5.3. Thermal noise method

The time average of the squared motion due to thermal noise $\langle x^2 \rangle$ has been used to obtain the TF effective elastic constant $k_{\text{eff}}$ following equation (9). A Stanford Research SR780 network signal analyzer has been used to measure the power spectral density for TFs A and B. An example of these spectra measured for a TF-A is shown in figure 5. The current noise background of 0.14 pA·Hz$^{-1/2}$ is due to the noise of the current to voltage amplifier. Once this background is subtracted the power spectral density can be integrated to obtain $\langle x^2 \rangle$. The effective elastic constants obtained from $k_{\text{eff}} = k_B T / \langle x^2 \rangle$ are $2045 \pm 102$ Nm$^{-1}$ for TF-A and $4220 \pm 215$ Nm$^{-1}$ for TF-B. Their elastic coupling constants $k_c$ obtained from equation (11) are $24 \pm 16$ Nm$^{-1}$ and $238 \pm 36$ Nm$^{-1}$ respectively. The thermal noise based method relies on the calibration of the oscillation amplitude from the current amplitude. As the mean squared amplitude is used to obtain the effective elastic constant, an error of a 5% in the oscillation amplitude calibration results in an error of 10% in the effective elastic constant.

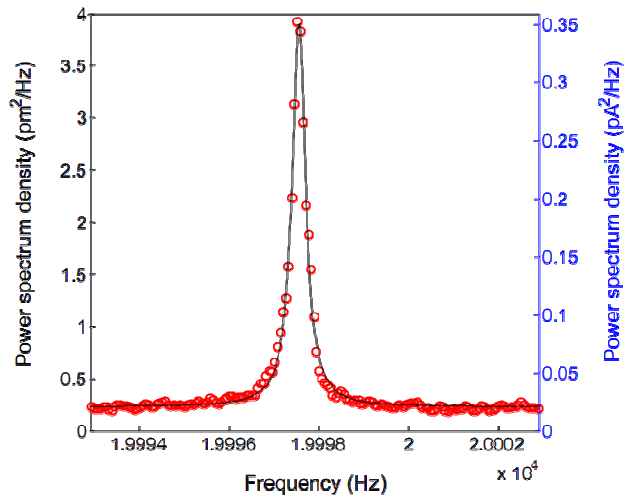

Figure 5. Power spectrum density of TF-A measured at room temperature in vacuum. The *Q* factor is about 55500 and the background current noise level due to the amplifier is $0.14 \text{ pA} \cdot \text{Hz}^{-1/2}$. The power spectrum density of the noise shows the expected lorentzian shape (solid line).

Table 2. Comparison of the obtained effective elastic constant $k_{\text{eff}}$, elastic constant of the coupling $k_c$ and sensitivity $\alpha \equiv \Delta f^{\text{anti-phase}} / \Delta k = f_0^{\text{anti-phase}} / 2k_{\text{eff}}$ following the three different approaches described in section 2.

|  | TF-A | | | TF-B | | |
| --- | --- | --- | --- | --- | --- | --- |
|  | $k_{\text{eff}}$ (Nm$^{-1}$) | $k_c$ (Nm$^{-1}$) | $\alpha$ (Hz·m·N$^{-1}$) | $k_{\text{eff}}$ (Nm$^{-1}$) | $k_c$ (Nm$^{-1}$) | $\alpha$ (Hz·m·N$^{-1}$) |
| Opto-mechanical | $2338 \pm 96$ | $98 \pm 4$ | $4.3 \pm 0.4$ | $4540 \pm 220$ | $318 \pm 15$ | $3.6 \pm 0.3$ |
| Cleveland variation | $2287 \pm 52$ | $85 \pm 15$ * | $4.4 \pm 0.2$ | $4505 \pm 234$ | $309 \pm 43$ ** | $3.6 \pm 0.4$ |
| Thermal noise | $2045 \pm 102$ | $24 \pm 16$ * | $4.9 \pm 0.5$ | $4220 \pm 215$ | $238 \pm 36$ ** | $3.9 \pm 0.4$ |

Values marked with * and ** have been obtained from equation (11) using $k = 974 \pm 40$ Nm$^{-1}$ and $k = 1634 \pm 79$ Nm$^{-1}$ respectively.

Table 2 shows the obtained values of effective elastic constant $k_{\text{eff}}$, elastic constant of the coupling $k_c$ and sensitivity $\alpha \equiv \Delta f^{\text{anti-phase}} / \Delta k = f_0^{\text{anti-phase}} / 2k_{\text{eff}}$ following the three different approaches presented in section 2. Although in reasonably good agreement, the effective elastic constants and elastic constants of the coupling obtained by thermal noise based method are slightly lower than the ones obtained with the opto-mechanical method or the variation of the Cleveland method. This is probably due to additional excitation sources. The relevance of the elastic constant of the coupling is clearly confirmed by the opto-mechanical method and the Cleveland variation method. When the coupling is neglected the effective elastic constant is underestimated by a factor about 15-35% depending on the TF studied. The effect of the coupling is stronger for TF-B. Note that a large number of TF sensors used in SPMs have similar geometrical dimensions and eigenfrequencies to those of TF-B and thus the effect of the coupling between the prongs should not be neglected.

### 6. Rebalancing the sensor

When a tip is attached to one prong of the TF the sensitivity $\alpha \equiv \Delta f^{\text{anti-phase}} / \Delta k$ is strongly reduced [25]. For example the reduction in $\alpha$ for a tip made of a tungsten wire 400 µm long and 50 µm in diameter can be about a factor 2 or 3. This can be counteracted by adding the same amount of mass to the other prong, i.e. rebalancing the TF. During this rebalancing process the $Q$ factor can rise up to values close to the $Q$ factor of the bare TF as previously reported in Ref. [7]. Thus, even though the calibration methods used in previous sections can be easily generalized to the case of a tip loaded TF it is desirable to rebalance the TF sensors to recover a high sensitivity $\alpha$ and a high $Q$ factor. In this context the variation of the Cleveland method [21] results very convenient. If the counter masses $\Delta m$ are attached to the tipless prong we can obtain the effective elastic constant $k_{\text{eff}}$ of the TF sensor and at the same time we recover a high sensitivity $\alpha$ and a high $Q$ factor. In order to attach the exact amount of counter mass to the tipless prong we proposed in a previous work [7] the use of the $Q$ factor of the anti-phase mode as an indicator of the balance ratio. This method is an iterative trial and error process and thus it can be time consuming. Here we propose to null the piezoelectric current from the in-phase mode instead of maximizing the $Q$ factor of the anti-phase mode. Electrodes in TFs are designed to suppress completely the in-phase mode current but when the masses of the prongs are different their oscillation amplitudes are also different and the suppression is not complete. The current from the in-phase mode of a TF-A has been measured together with the oscillation amplitude of both prongs while test masses were attached or detached to one prong (Figure 6). We have found a linear relation between the in-phase mode current and the mass difference of the prongs making possible to rebalance the TF in 2-3 iterations.

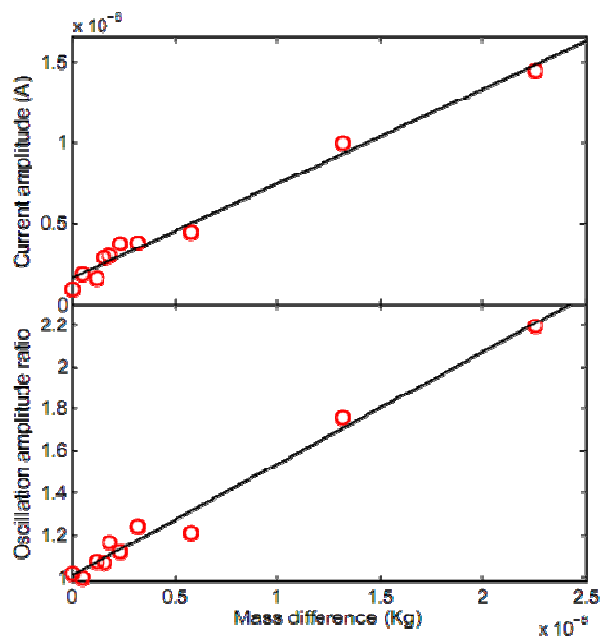

Final draft post-refereeing:
A Castellanos-Gomez *et al* 2009 *Nanotechnology* 20 215502.
doi: 10.1088/0957-4484/20/21/215502Figure 6. (Top panel) Measured in-phase mode current amplitude as a function of the mass difference between the prongs of TF-A. (Bottom panel) Simultaneously, the oscillation amplitude ratio has been optically measured. For a negligible mass difference the oscillation amplitude of both prongs is the same and the current of the in-phase mode is almost suppressed. While the oscillation amplitude ratio is close to 1 for a balanced TF, the current of the in-phase mode is finite. This can be attributed to small differences in the electrode configuration of each prong.

## 7. Conclusions

We have experimentally characterized the dynamics of TFs sensors. We show that a coupled harmonic oscillators model which includes a finite coupling between the prongs is in remarkable agreement with the observed motion of TFs. Furthermore the commonly used single harmonic oscillator model is not valid because of the crucial role of a non negligible coupling between the TF prongs. We have proposed three different experimental procedures to determine the elastic constant of the coupling between the prongs: an opto-mechanical method, a variation of the Cleveland method and a thermal noise based method. The results show that a weakly coupled oscillators approximation is inaccurate for commercially available TF used in SPMs. The precise determination of the elastic coupling between the prongs of a TF allows to obtain a quantitative relation between the frequency shift and the force gradient acting at the free end of a TF prong.

**Acknowledgements**

A.C-G. acknowledges fellowship support from the Comunidad de Madrid (Spain). This work was supported by MICINN (Spain) (MAT2008-01735 and CONSOLIDER-INGENIO-2010 CSD-2007-00010) and Comunidad de Madrid (Spain) through the program Citecnomik (S_0505/ESP/0337).